\documentclass[twocolumn]{aastex631}

\usepackage{chngpage}
\usepackage{booktabs}
\usepackage{amsmath}

\newcommand{\uvec}[1]{\boldsymbol{\hat{\textbf{\textit{#1}}}}}

\newcommand\RGMX{\bgroup\markoverwith{\textcolor{cyan}{\rule[0.5ex]{4pt}{1pt}}}\ULon}

\newcommand\ACCX{\bgroup\markoverwith{\textcolor{red}{\rule[0.5ex]{4pt}{1pt}}}\ULon}
\shorttitle{CBP formation in misaligned disks}
\shortauthors{Childs \& Martin}
\graphicspath{{./}{figures/}}

\begin{document}

\title{Misalignment of terrestrial circumbinary planets as an indicator of their formation mechanism}

\author[0000-0002-9343-8612]{Anna C. Childs}
\author[0000-0003-2401-7168]{Rebecca G. Martin}
\affiliation{Nevada Center for Astrophysics, University of Nevada, Las Vegas, NV 89154, USA}
\affiliation{Department of Physics and Astronomy, University of Nevada, Las Vegas, 4505 South Maryland Parkway,
Las Vegas, NV 89154, USA}



\begin{abstract}
Circumbinary gas disks are often observed to be misaligned to the binary orbit suggesting that planet formation may proceed in a misaligned disk. With $n$-body simulations  we consider the formation of circumbinary terrestrial planets from a particle disk that is initially misaligned.  We find that if terrestrial planets form in this way, in the absence of gas, they can only form close to coplanar or close to polar to the binary orbit. Planets around a circular binary form coplanar while planets around an eccentric binary can form coplanar or polar depending on the initial disk misalignment and the binary eccentricity.  The more massive a terrestrial planet is, the more aligned it is (to coplanar or polar) since it has undergone more mergers that lead on average to smaller misalignment angles. Nodal precession of particle disks with very large initial inclinations lead to high mutual inclinations between the particles. This produces high relative velocities between particles that leads to mass ejections that can completely inhibit planet formation.    Misaligned terrestrial circumbinary planets may be able to form in the presence of a misaligned circumbinary gas   disk that may help to nodally align the particle orbits and maintain the inclination of the planets during their formation. 
\end{abstract}

\keywords{Binary stars (154), Solar system terrestrial planets (797), Extrasolar rocky planets (511), Planet formation (1241)}

\section{Introduction} \label{sec:intro}

Circumbinary gas disks around young stars are commonly observed to be misaligned to the binary orbital plane. The degree of misalignment often increases with binary separation and eccentricity \citep{Czekala2019}. Misalignments in the range $25^{\circ}-60^{\circ}$ have been observed around GG Tau A \citep{Kohler2011, Andrews2014}, KH 15D \citep{Chiang2004, Winn2004,Fang2019}, IRS 43 \citep{Brinch2016}, and L1551 NE \citep{Takakuwa2017} while the gas disk around HD 98800 B and debris disk around 99 Herculis are misaligned by almost $90^{\circ}$ \citep{Kennedy2012, Kennedy2019, Smallwood2020}.  A disk (or planet) that is misaligned by $90^{\circ}$ with an angular momentum vector that is parallel to the binary eccentricity vector is referred to as being polar-aligned and this is a stable configuration \citep{Martin2017,Lubow2018, Zanazzi2018,Cuello2019,Chen2020}.   

The chaotic gravitational collapse of a molecular cloud to form a binary star system and a subsequent circumbinary disk may result in the misalignment.  Misalignment may occur as a result of turbulence in the molecular gas cloud \citep{Offner2010, Tokuda2014, Bate2012}, later accretion of material by the young binary \citep{Bates2010, Bate2018}, warping by a tertiary companion or stellar flyby \citep{Nealon2020} or, if the binary forms from a cloud whose elongated axis is misaligned to its rotation axis \citep{Bonnell1992}. 

Circumbinary planet (CBP) transit timing variations are much larger than the duration of the transit, due to the motion of the binary, which makes CBPs difficult to detect.  As a result, most CBP transits are detected by eye which has a strong observational bias against small planets with small transit depths.  While algorithms for automated detection of coplanar CBPs have been presented, terrestrial CBPs have yet to be observed \citep{Windemuth, MartinFabrycky2021}.
The difficulty of transit detection for CBP's increases with planet misalignment \citep{Schneider,MartinD2014, Martin_2017}.  The number of transits per epoch for misaligned planets is atypical, thus enhancing the difficult of photometric detection  \citep{Chen2021}.

All the so-far observed CBPs are nearly coplanar. The most highly misaligned CBPs that have been observed are Kepler-413b and Kepler-453b, and they have modest inclinations of about $2.5^{\circ}$ to the binary orbital plane \citep{Kostov2014, Welsh2015}.  While \cite{Armstrong2014} presented occurrence rates for CBPs by assuming that CBPs either preferentially form coplanar or there is an isotropic distribution of misaligned CBPs, \cite{Gongjie2016} argued that the lack of observed misaligned planets is representative of nature. In support of this view, the BEBOP radial-velocity survey for circumbinary planets has constrained the distribution of CBP inclinations to be $<10^{\circ}$ \citep{Martin2019}. However, misaligned planets may be more likely around wider binaries with orbital periods ranging from $30-10^5 \, \rm days$ \citep{Czekala2019}.

The standard model of the late stage of terrestrial planet formation refers to in situ planet growth in a gas free environment via core accretion \citep{Artymowicz1987, Lissauer1993, Pollack1996}. It is still debated whether planets form in the presence of gas and experience inward migration to their observed orbital periods or, if planets form in situ \citep{Bromley2015, Penzlin2020}.    Each scenario has its own barriers.  If coplanar CBPs form in the presence of gas, they may migrate through a series of unstable resonant locations that is likely to lead to ejection for small planets in highly turbulent disks \citep{Martin2022}.   The resonance strength decreases with inclination relative to the binary \citep{Lubow2018}.  Previous studies have shown that in the absence of gas, CBPs may form in situ from a coplanar particle disk \citep{Quintana2006,Childs2021} or from a polar particle disk \citep{Childs2021ApJ}. However, there remain uncertainties over how planetesimals are able to form in situ.

In this letter, for the first time, we investigate the late stage of in situ terrestrial planet formation in a
disk that is highly misaligned to one of these two stationary states.
\cite{Zhang2018} found that terrestrial planets in an initially misaligned particle disk around one component of a binary can form preferentially coplanar, depending on the binary separation.
We use $n$-body simulations to model the late stage of terrestrial planet formation around circular and eccentric binaries in an initially highly misaligned circumbinary disk.    In Section 2 we discuss the setup of our simulations. In Section 3 we present our results, and in Section 4 we conclude with a summary of our findings.

\section{Simulations}

In this section we describe our simulations of the late stage of terrestrial CBP formation around a misaligned binary star. We use  the $n$-body code \texttt{REBOUND} and high precision integrator \texttt{IAS15} \citep{Rein2012, Rein2015}.

\begin{table*}
\caption{Each row describes the model name, binary eccentricity ($e_{\rm b}$), initial inclination ($i_{\rm b0}$), and initial particle surface density fit ($\Sigma$) \citep[from Figure 2 in][]{Childs2021ApJ}.  Columns 5-14 show the average values and standard deviations for the terrestrial planet multiplicity (\#), planet mass ($M_{\rm p}$), semi-major axis ($a_{\rm p}$), eccentricity ($e$) and misalignment ($i_{\rm b}$ for C30 and C60 and $i_{e}$  for P60) after $7\,\rm Myr$ of integration time.  The two sets of  statistics consider bodies with a mass  $M_{\rm p}\ge 0.1\,M_{\oplus}$ and bodies with mass  $M_{\rm p}\ge 1.0\,M_{\oplus}$.  The last column shows the total mass ejected ($M_{\rm e}$)
normalized by the initial disk mass ($M_{\rm d}$).
} 
\begin{adjustwidth}[]{-1in}{}
\resizebox{0.9\linewidth}{!}{
\hskip-4.0cm
\begin{tabular}{cccc|ccccc|ccccc|c}
    \hline
  {} & {} & {} & {} & \multicolumn{5}{c|}{$M_{\rm p}/M_\oplus \ge 0.1$}  &  \multicolumn{5}{c|}{$ M_{\rm p}/M_\oplus \ge 1.0$} & {}   \\
  \hline
{Model}  & {$e_{\rm b}$} & {$i_{\rm b0}^{\circ}$} & {$\Sigma$} & 
{\#} &  {$M_{\rm p}/M_\oplus$} &  {$a_{\rm p}/ \rm au$} &  {$e$} &  {$i_{e/ \rm b}^{\circ}$} & 
{\#} &  {$M_{\rm p}/M_\oplus$} &  {$a_{\rm p}/ \rm au$} &  {$e$} &  {$i_{e/\rm b}^{\circ}$} & {$M_{\rm e}/M_{\rm d}$}  \\
\hline
C30 & 0.0  & 30.0 & CC & $4.8 \pm 1.3$ & $0.9 \pm 0.8$ & $2.6 \pm 0.8$ & $0.07 \pm 0.06$ & $12.6 \pm 8.1$ &

$1.9 \pm 0.5$ &$1.8 \pm 0.5$ &$2.2 \pm 0.3$ & $0.06 \pm 0.03$ & $6.9 \pm 2.9 $ &$0.06 \pm 0.02$ \\

C60 & 0.0  & 60.0 & CC & $1.5 \pm 0.9$ & $0.3 \pm 0.2$ & $2.4 \pm 0.7$ & $0.16 \pm 0.10$ & $51.5 \pm 5.5$ &

$0.02 \pm 0.13$ &$1.1 \pm 0.0$ &$1.5 \pm 0.0$& $0.02 \pm 0.0$ & $44.7 \pm 0.0$ & $0.84 \pm 0.07$ \\

P60 & 0.8 & 60.0 & EP & $3.7 \pm 1.0$ & $1.1 \pm 0.9$ & $2.5 \pm 0.7$ & $0.07 \pm 0.04$ & $13.4 \pm 8.6$ &$1.8 \pm 0.5$& $1.9 \pm 0.5$ & $2.3 \pm 0.4$ & $0.06 \pm 0.04$ & $7.0 \pm 3.3$ & $0.13 \pm 0.05$ \\

\hline
C30$_{\rm JS}$ & 0.0   & 30.0 & CC & $2.0 \pm 0.7$ & $1.5 \pm 1.0$ & $2.0 \pm 0.5$ & $0.08 \pm 0.06$ & $9.0 \pm 5.8$  &$1.2 \pm 0.4$&$2.2\pm 0.6$&$2.0 \pm 0.2$&$0.07\pm0.05$&$6.9\pm 3.4$& $0.38 \pm 0.09$ \\

C60$_{\rm JS}$ & 0.0  & 60.0 & CC & $0.3 \pm 0.5$ & $0.2 \pm 0.2$ & $2.5 \pm 3.4$ & $0.15 \pm 0.09$ & $53.4 \pm 7.6$ 

&$0.0 \pm 0.0$&-&-&-&-& $0.96 \pm 0.03$ \\

P60$_{\rm JS}$ & 0.8 & 60.0 & EP & $1.1 \pm 0.8$ & $1.1 \pm 0.6$ & $2.0 \pm 0.3$ & $0.08 \pm 0.04$ & $11.1 \pm 6.6$&$0.7 \pm 0.5$&$1.6 \pm 0.3$&$1.9 \pm 0.1$&$0.08 \pm 0.04$&$8.8 \pm 4.6$& $0.73 \pm 0.16$\\

 \hline
\end{tabular}
}
\end{adjustwidth}
\label{tab:avg_planets}
\end{table*}

\begin{figure*}
\centering
		\includegraphics[width=1\columnwidth,height=0.5\textheight]{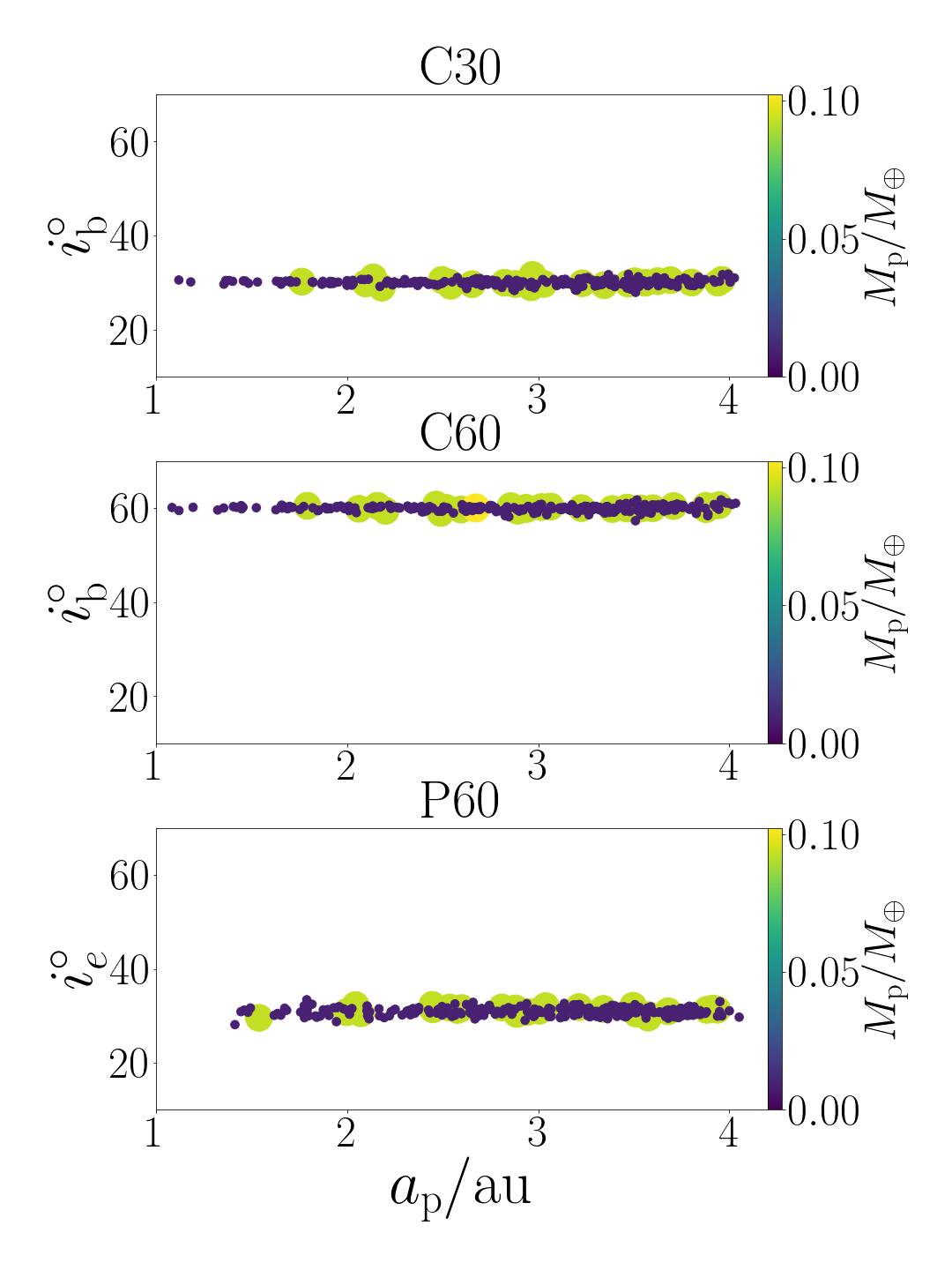}
			\includegraphics[width=1\columnwidth,height=0.5\textheight]{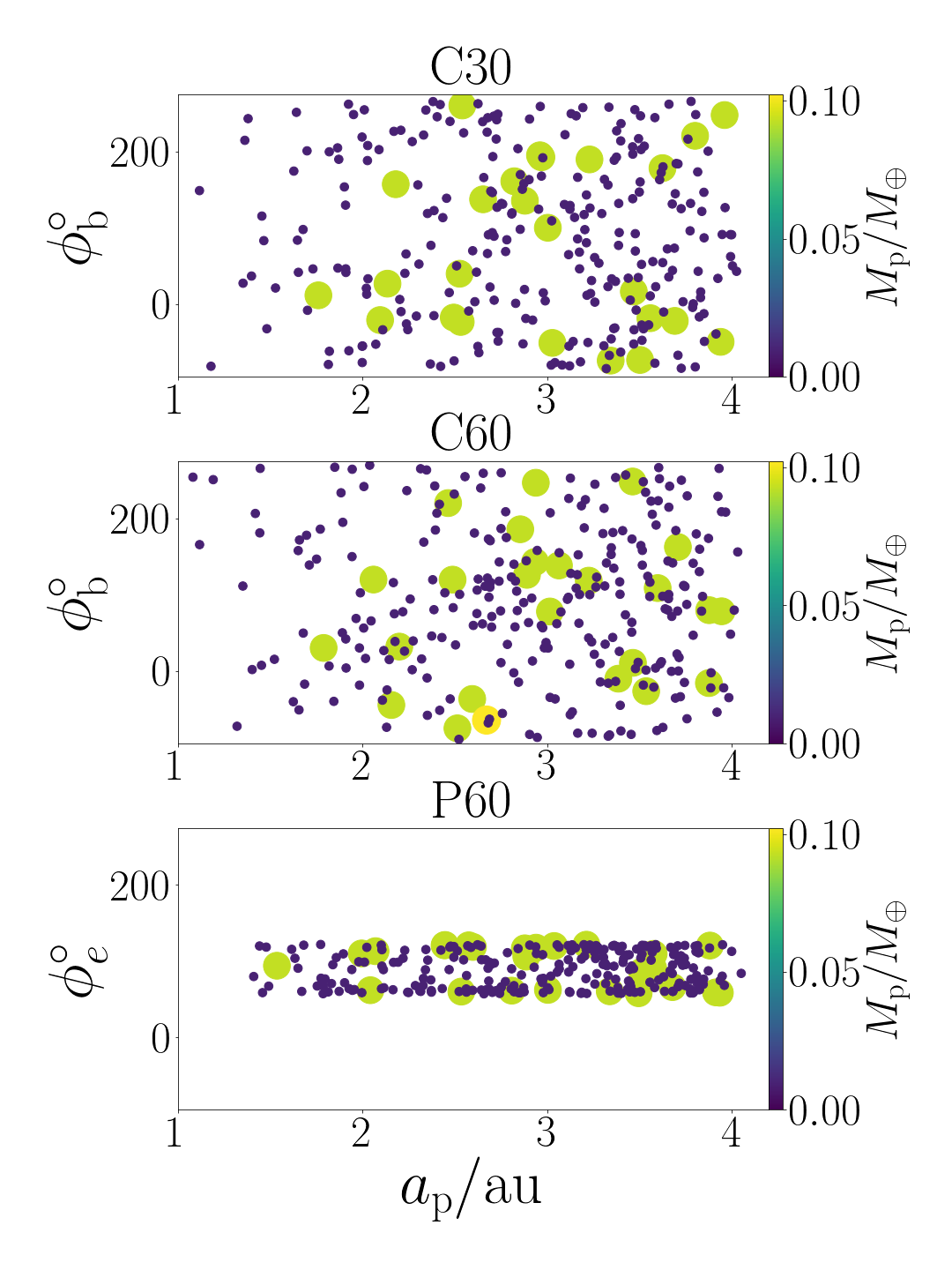}
    \caption{Particle misalignments (left) and nodal phase angles (right) as a function of semi-major axis at a time of $100\,\rm Kyr$ of integration time with no expansion factor in the systems without Jupiter and Saturn.  The size and color of the points are correlated to the body mass.    There is not much change to the body masses (which begin at either $\approx 0.1 \, M_{\oplus}$ or $\approx 0.01 \, M_{\oplus}$) or inclinations at this time.}
    \label{fig:start_phases}
\end{figure*}

\subsection{Initial particle disk}

The binary stars have equal masses $M_1=M_2=0.5 \, M$, where the total mass is $M=1.0 \, M_{\odot}$. The semi-major axis is $a_{\rm b}=0.5 \, \rm au$ and we consider two different binary eccentricities of $e_{\rm b}=0.0$ and $e_{\rm b}=0.8$.  Initially we incline  the binary orbital plane relative to the particle disk by inclination angle $i_{\rm b0}$. We consider two different initial inclinations for the circular orbit binary of $i_{\rm b0}=30^\circ$ (C30) and $i_{\rm b0}=60^\circ$ (C60) and one inclination for the eccentric binary of $i_{\rm b0}=60^\circ$ (P60).

The particle disk begins with 260 Moon-sized planetesimals and 26 Mars-sized embryos on nearly circular and coplanar orbits which marks the onset of the late stage of terrestrial planet formation \citep{Weidenschilling77, Rafikov_2003}.  This mass distribution is successful at reproducing the masses of the solar system terrestrial planets and so it is commonly used in $n$-body simulations of terrestrial planet formation \citep{KOKUBO1996, Chambers2001, Quintana2014, Childs19}.    Circumbinary disks may have difficulty growing planetesimals in the inner regions due to tidal forces from the binary which inhibit pebble accretion and in situ planetesimal formation \citep{Moriwaki2004, Scholl2007, Marzari2013, Rafikov2015, Paardekooper_2012}.  Planetesimals experience high relative velocities in the inner disk and are more likely to fragment than accrete one another.  However, if fragmentation is accounted for, second generation planetesimals may be able to grow from the fragments and  kilometer-sized plantesimals may grow by two orders of magnitude in size \citep{Paardekooper2010}.    These problems may not be so severe in an misaligned or polar disc where the tidal forces are weaker \citep{Lubow2018}. If the barriers of planetesimal formation are able to be overcome, we expect that the early mass distributions for terrestrial planet formation will be similar to the circumstellar case.

All particle eccentricities are uniformly distributed in the range (0.0,0.01) and begin on nearly coplanar orbits with inclinations uniformly distributed between ($0^{\circ},1^{\circ}$).  The longitude of ascending node, argument of pericenter, and true anomaly are uniformly distributed between $0^{\circ}$ and $360^{\circ}$. All the particles are spherical and are given an initial density of $3\,\rm  g\,cm^{-3}$.  

The particles are initially distributed between semi-major axes of $a_{\rm p}=1-4 \, \rm au$.  The surface density profiles ($\Sigma$) we use for each particle disk setup are motivated by Smooth Particle Hydrodynamic (SPH) gas disk simulations around an equal mass binary.  The SPH data and analytic fits used for $\Sigma$ may be seen in Figure 2 in \citet{Childs2021ApJ}. The surface density profiles are for a steady state gas disk around a circular coplanar (CC) and an eccentric polar (EP) binary. 
If solid bodies grow quickly in a gas disk, planet-disk interactions can alter the distribution of the solid bodies. If the bodies do not grow substantially large in the gas disk and the gas disk dissipates quickly, the gas profile is representative of the starting distributions for the late stage of terrestrial planet formation.    The gas disk dissipation timescale may be as short as $100 \, \rm Kyr$ in circumstellar disks where photoevaporation is efficient \citep{Clarke2001}. Circumbinary disks are expected to have a similarly short dissipation timescale \citep{Alexander2012,Owen2012}.



Table \ref{tab:avg_planets} lists the binary parameters and the associated model name for each setup. To understand the effects of giant planets on terrestrial planet formation in misaligned disks, we add Jupiter and Saturn at their current mass and orbit in half of our simulations.  Runs that include Jupiter and Saturn have a ``JS" in the model name.  
Each model has 50 runs where the random seed number, for assigning the randomly chosen orbital elements of the particles, is changed in each run.  We 
integrate each run for a total time  of $7 \, \rm Myr$.

\subsection{Analysis of the particle orbits}

 We analyze the orbits of the bodies in the frame of the binary. 
We define $\textbf{\textit{r}}_{\rm b}$ as the instantaneous position vector of the binary,
$\textbf{\textit{e}}_{\rm b}$ as the instantaneous eccentricity vector of the binary, $\textbf{\textit{l}}_{\rm b}$ as the instantaneous angular momentum vector of the binary, and $\textbf{\textit{l}}_{\rm p}$ as the instantaneous angular momentum vector of the particle relative to the binary.

For a circular orbit binary (models C30 and C60), the misalignment is the inclination of the particle orbit  relative to the binary angular momentum vector given by
\begin{equation}
    i_{\rm b} =\textrm{cos}^{-1} (\uvec{l}_{\rm b} \cdot \uvec{l}_{\rm p}),
\end{equation}
where $\uvec{}$ denotes a unit vector,
and the nodal phase angle of the particle is calculated with
\begin{equation}
    \phi_{b} = \textrm{tan}^{-1} \left( \frac{\uvec{l}_{\rm p}\cdot (\uvec{l}_{\rm b} \times \uvec{r}_{\rm b})}{\uvec{l}_{\rm p} \cdot \uvec{r}_{\rm b}} \right ) + \frac{\pi}{2}.
\end{equation}
For the systems around an eccentric binary (P60), we calculate the misalignment as the inclination of a particle orbit relative to the binary eccentricity vector with
\begin{equation}
    i_{\rm e}=\textrm{cos}^{-1} (\uvec{e}_{\rm b} \cdot \uvec{l}_{\rm p}),
\end{equation}
and the nodal phase angle of the particle with
\begin{equation}
    \phi_{\rm e} = \textrm{tan}^{-1} \left( \frac{\uvec{l}_{\rm p}\cdot (\uvec{l}_{\rm b} \times \uvec{e}_{\rm b})}{\uvec{l}_{\rm p} \cdot \uvec{e}_{\rm b}} \right ) + \frac{\pi}{2}
\end{equation}
 \citep{Chen2019}. 

\subsection{Nodal precession}

A test particle that is coplanar ($i_{\rm b}=0^\circ$) to a binary orbit or polar to an eccentric orbit binary ($i_{\rm b}=90^\circ$ and $\phi_{\rm e}=90^\circ$ or alternatively $i_{\rm e}=0^\circ$) is in a stationary orbit \citep{Farago2010, Chen2019}. A particle that is at any other inclination  is misaligned and undergoes nodal preccesion due to the gravitational torque  from the binary.

The nodal precession of a circumbinary particle undergoes either circulation or libration   \citep{Doolin2011}. In a circulating orbit, the angular momentum vector of the particle precesses around the binary angular momentum vector and the ascending node circulates over $360^{\circ}$. Around a circular orbit binary, the particle always circulates. In a librating orbit, the angular momentum vector of the particle precesses about the binary eccentricity vector and the ascending node oscillates over only a limited range of angles.  Around an eccentric binary, a particle can librate or circulate depending on its initial inclination and the eccentricity of the binary.  The critical inclination that separates the librating and circulating orbits for a test particles is
\begin{equation}
    i_{\rm crit} = \rm{sin^{-1}} \sqrt{\frac{1-\textit{e}_{\rm b}^2}{1+4\textit{e}_{\rm b}^2}}
\end{equation}
\citep{Farago2010}.  The critical inclination for a binary with  $e_{\rm b}=0.8$ is $i_{\rm crit}=18.5^{\circ}$.  The initial misalignment we consider around the eccentric binary is above this critical angle and therefore the particles are all in librating orbits initially in model P60.

The nodal precession frequency for a particle at radius $a_{\rm p}$ is given by
\begin{equation}
    \omega =  \frac{3}{4} k \frac{M_1 M_2}{M^2} \left ( \frac{a_{\rm b}}{a_{\rm p}} \right )^{7/2} \Omega_{\rm b},
\end{equation}
where the angular frequency of the binary is
\begin{equation}
    \Omega_{\rm b}=\sqrt{\frac{GM}{a_{\rm b}^3}}.
\end{equation}
If the particle is close to a polar alignment, it precesses about the binary eccentricity vector with
\begin{equation}
k=3 \sqrt{5}e_{\rm b} \sqrt{1+4e_{\rm b}^2}
\end{equation}
\citep{Farago2010,Lubow2018}. If the particle is close to coplanar, it precesses about the binary angular momentum vector with
\begin{equation}
    k = \sqrt{1+3e_{\rm b}^2-4e_{\rm b}^4}
\end{equation}
 \citep{Smallwood2019}.  
The precession period associated with each nodal precession frequency is
\begin{equation}
    P = \frac{2 \pi}{\omega}.
\end{equation}
The precession period at the outer edge of our particle disk, $r=4 \, \rm au$, is $\sim 0.8 \, \rm Kyr$ around the eccentric binary (for P60) and $\sim 2.7 \, \rm Kyr$  around the circular binary (for C30 and C60).

\begin{figure*}
\centering
	\includegraphics[width=2\columnwidth,height=0.35\textheight]{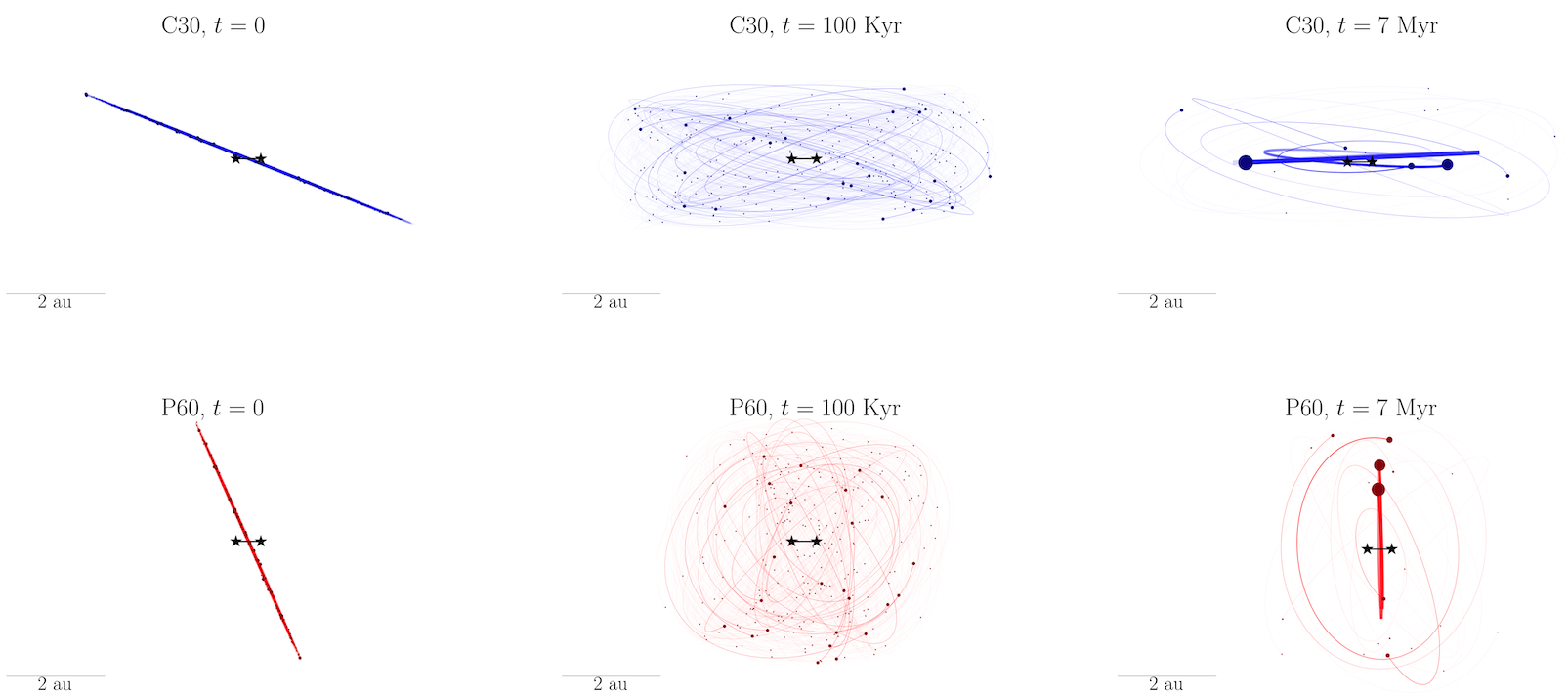} 
    \caption{  Particle and binary orbits of C30 (upper panels, circular orbit binary with $i_{\rm b 0}=30^\circ$) and P60 (lower panels, eccentric orbit binary with $i_{\rm b 0}=60^\circ$) at three different times: $t=0$, $100 \, \rm Kyr,$ and $7 \, \rm Myr$.}  The binary is shown with black stars and its orbit is shown edge on.
    \label{fig:setups}
\end{figure*}

\subsection{Expansion factor}

We use the non-symplectic integrator $\texttt {IAS15}$ in order to accurately simulate the motion of the binary and its effects on the circumbinary disk.  This integrator utilizes an adaptive time step that is highly precise but comes at a large CPU cost.  In order to reduce the computation time, we make use of an expansion factor \citep{KOKUBO1996, Kokubo}.  After an initial relaxation time of $100\,\rm Kyr$ (well in excess of the nodal precession timescales for all particles), we artificially inflate the radii of the particles by a factor $f=25$.  Doing so decreases the collision time which, when collisions are modeled as only perfect mergers, quickly decreases the number of bodies in the system and reduces the computation time. 
Although a large expansion factor may lead to somewhat different system architectures, the general planet formation trends that emerge are similar to those found in simulations with lower values of $f$ \citep{Childs2021, Childs2022}.


Fig.~\ref{fig:start_phases} displays the particle inclinations and nodal phase angles at a time of $100 \, \rm Kyr$ for all systems without giant planets.  We find similar distributions in our systems with giant planets at $100 \, \rm Kyr$ although some of the particles close to the giant planets are already ejected.
At this time, the particles have become randomized in their nodal phase angle with inclinations that remain mostly unchanged from their initial values.    The bodies remain highly inclined since they have not yet interacted with one another.  The particles around the circular orbit binary are in circulating orbits and so their nodal phase angles range from $0-360^\circ$ while the particles around the eccentric binary are in librating orbits and so their nodal phase angles have a limited range of values centered on $\phi_{\rm e}=90^\circ$. There is little evolution of the particle semi-major axes and inclinations within the first $100 \rm \, Kyr$.  As a result, there is only one collision in the C60 system where a planetesimal merged with an embryo.  Besides this one embryo, the particle masses have not changed from their original masses within this time. 

  We note that our initial condition of a flat but tilted particle disk with no gas is somewhat idealised. The particles may be aligned to the gas disk while it is massive, but as it dissipates, particles may evolve separately to the gas.  Given that very few collisions have occured in the first $100\,\rm Kyr$ of evolution with $f=1$, we see that the collision timescale for the particles is much longer than the gas disk dissipation timescale.  Therefore even if different parts of the disc disperse at different times, we do not expect this to change our results significantly because there will be very few collisions during the period of time where there is a partial gas disk.  Understanding the effects of the partial gas disk on the solid bodies is outside of the scope of this paper, but these effects are likely secondary to gravitational interactions and we do not expect it to significantly affect the final planetary system.


Fig. \ref{fig:setups} shows the particle and binary orbits for an example simulation in the C30 and P60 cases. The left panels show the initially flat but tilted particle disk. The binary eccentricity vector  is in the $x$-direction. The middle panels show the disk at a time of $100\,\rm Kyr$ after evolution with $f=1$. The particle disk forms a thick annulus as the nodal phase angles have become randomly distributed but the inclinations remain constant.  After this initial phase of evolution, we then employ an expansion factor of $f=25$ of the particle radii and continue integrating the systems for $7 \, \rm Myr$ total of simulation time. We describe the outcomes of these simulations in the next section and show an example of the final systems in the rightmost panels of Fig. \ref{fig:setups}.

\section{Results}

\begin{figure*}
\centering
	\includegraphics[width=0.9\columnwidth]{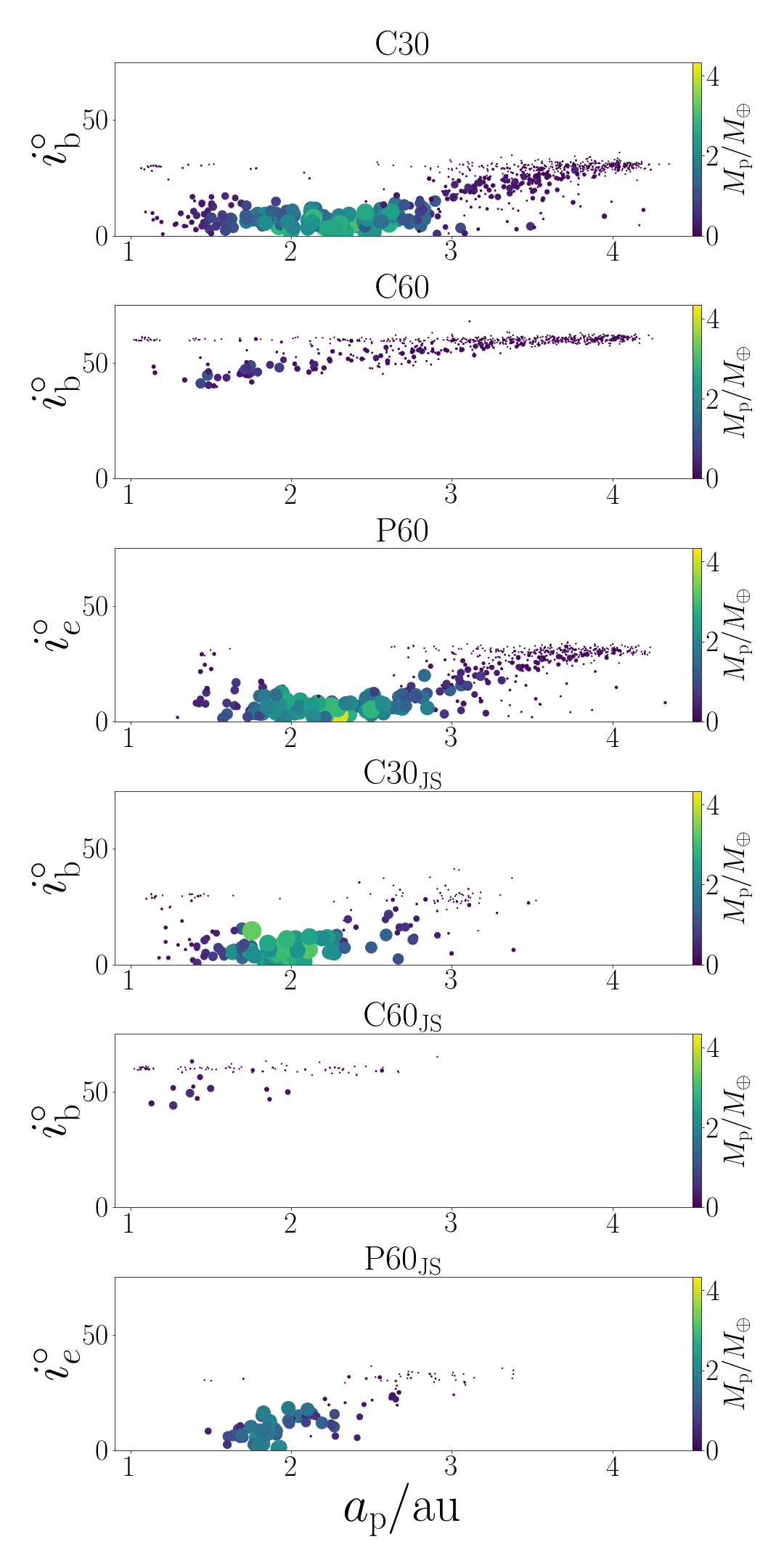}
		\includegraphics[width=0.9\columnwidth]{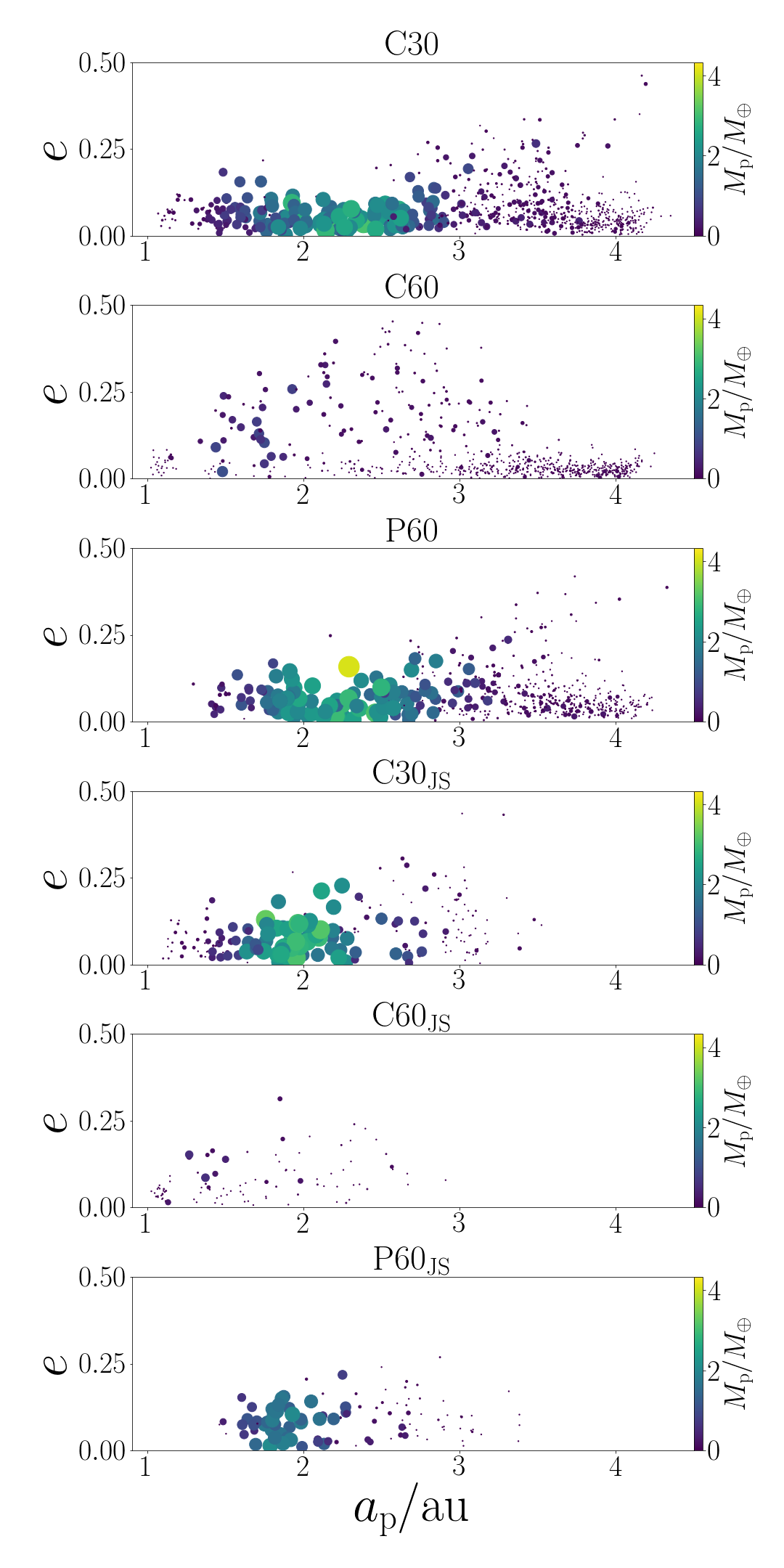}
    \caption{Misalignment (left panels, $i_{\rm b}$ for C30 and C60 and $i_{e}$ for P60) and eccentricity (right panels) versus semi-major axis for all the remaining bodies at time $t=7 \, \rm Myr$.  The size and color of the points are correlated to the body mass.}
    \label{fig:inc_v_a}
\end{figure*}

In columns 5-16 in Table \ref{tab:avg_planets} we list the results of the simulations at time $t=7\,\rm Myr$.
We list the average values for the planet multiplicity (\#), planet mass ($M_{\rm p}$), semi-major axis ($a_{\rm p}$), eccentricity ($e$), and inclination ($i^{\circ}$). We consider two mass ranges, the first is for bodies with $M_{\rm p} \ge 0.1\, M_{\oplus}$
The second is
for bodies with $M_{\rm p} \ge 1.0 \, M_{\oplus}$. 
The error bars represent the standard deviation.  The last column shows the mass ejected from the system ($M_{\rm e}$) 
normalized by the initial particle disk mass ($M_{\rm d}$). 
Fig. \ref{fig:inc_v_a} shows the misalignment versus semi-major axis on the left and eccentricity versus semi-major axis on the right, for all the bodies, across all 50 runs, at $7 \, \rm Myr$ in a given system. 


The C30 and P60 systems are both initially misaligned by $30^{\circ}$ to a stationary alignment and they form very similar planetary systems with a relatively small fraction of the mass ejected. The average inclination of the planets that form are $\lesssim 10^\circ$ from coplanar (C30) or polar (P60). Therefore, planets that form from a misaligned circumbinary disk, preferentially form either close to coplanar or close to polar.  


The more massive a terrestrial planet is, the more mergers that have taken place during its formation and the less misaligned it is on average. This is because collisions between bodies with random nodal phase angles lead to a decrease in the average inclination of the bodies. Fig.~\ref{fig:inc_v_a} shows that in C30 and P60, the misalignment increases with separation while the planet size   first increases then decreases.  The planets close in remain small because of the low surface density there. Farther out, the particles are more widely spaced and the collision timescale is longer. There hasn't been enough time yet for these outer bodies to merge into planets.    The most misaligned bodies left in these systems are the planetesimals and embryos that did not interact with the other bodies and they remain at their initial inclination.


The C60 system (that is inclined by $60^\circ$ to a stationary alignment) loses $84\%$ of the disk mass through ejections, which completely inhibits planet growth. In all simulations, mass ejection dominates stellar collision \citep[see also][]{Smullen2016}. The high mutual inclinations between the bodies in C60 leads to very high relative velocities. Mass ejections generally take place shortly after a merger.   As a result, $90 \%$ of the collisions in the C60 runs result in high energy impacts that place the post-collision body on a hyperbolic orbit.   These systems are also the only systems to lose mass to the binary.  We extended the C60 runs to $20 \, \rm Myr$ 
and found that $\sim 50\%$ of the bodies remaining at $7 \, \rm Myr$ are ejected from the system by this time.  We expect similar results to C60 if we were to consider a disk inclined by $i_{b0}=30^\circ$ around the eccentric binary since the orbits would be librating and misaligned by $60^\circ$ to the stationary polar inclination.

  Our use of an expansion factor may lead to lower particle eccentricities since bodies can merge before their orbits have time to grow to more excited states. If the particles were allowed to grow to higher eccentricity, we would expect higher mass loss rates.  Large mutual inclinations between particles also cause high relative velocities in our simulations. Because we ran the simulations initially with $f=1$, the mutual inclinations are not affected by our large expansion factor.



When giant planets are added to a system, they further destabilize regions in space which leads to higher ejection rates. However, the perturbations from the exterior giant planets also leads to higher collision rates between the inner terrestrial bodies.  As a result, we observe fewer but more massive planets in all the systems with Jupiter and Saturn.  We also observe that the terrestrial planets form closer-in to the binary with generally, more eccentric and less misaligned planets.


The giant planets undergo inclination oscillations in all systems, as a result of planet-planet interactions around the circular orbit binary and both planet-planet and planet-binary interactions around the eccentric binary \citep{Chen2022}.  Jupiter is able to remain highly inclined in all systems, except in one run in C$60_{\rm JS}$.  After undergoing large oscillations, Saturn is ejected from all C$60_{\rm JS}$ runs, around $2 \, \rm Myr$.  
Saturn is ejected from $26\%$ of the P$60_{\rm JS}$ runs.  No giant planets are ejected from any of the C$30_{\rm JS}$ runs.

\section{Conclusions}

We have explored the formation of terrestrial planets from a circumbinary particle disk that is misaligned relative to the binary orbital plane.  We consider the late stage of planet formation where planets form in situ, via core accretion, immediately after the dissipation of a misaligned gas disk.  Despite the initial disk inclination, we found that if terrestrial planets form, they are  close to coplanar to a circular binary and close to coplanar or polar to an eccentric binary depending on the initial disk inclination and the binary eccentricity. The more massive a terrestrial planet is, the less misaligned it will be (from coplanar or polar) as it has undergone more mergers that reduce the average inclination. However, if the initial particle disk misalignment is too high, particles can have high relative velocities which lead to particle-particle scattering that inhibits planet formation.    Small bodies may survive closer in to the binary but the more massive terrestrial planets are found further out where the surface density is higher and collisions are more likely to take place.  In the future, if Earth-sized terrestrial CBPs are observed in small orbits, close to the stability limit, we suggest that they may have migrated there.


Our findings show that core accretion in the late stage of terrestrial planet formation is only able to produce coplanar or polar terrestrial CBPs in the absence of gas. If  highly misaligned terrestrial CBPs are observed in the future,   we suggest that they cannot have formed in this way. One possibility for their formation is 
in a   misaligned circumbinary gas disk. In this scenario,   the planet may remain coplanar to the gas disk as it undergoes nodal precession \citep[e.g.][]{Lubow2016}.
This requires the planet mass to be sufficiently small that it does not open a gap in the gas disk. Once a gap is opened in a circumbinary disk, planet-disk interactions may lead to lower levels of misalignment \citep{Pierens2018}. Planet-planet interactions could help maintain  the misalignment after gas dissipation \citep{Chen2022}   although for very close binaries, stellar tides may realign a highly inclined planet on Gyr timescales \citep{Correia2016}.  Our simulations including giant planets show that misaligned exterior giant planets do not have a significant effect on the orbits of the inner aligned terrestrial planets.

\begin{acknowledgements}
We are very grateful to David Martin for contributing his expertise and useful comments that improved the manuscript.  We acknowledge support from NASA through grant 80NSSC21K0395.  AC acknowledges support from a UNLV graduate assistantship.  Computer support was provided by UNLV’s National Supercomputing Center.

\end{acknowledgements}

\bibliography{ref}{}
\bibliographystyle{aasjournal}



\end{document}